\documentclass[floatfix,aps,prl,twocolumn,superscriptaddress]{revtex4-1}

\pdfoutput=1
\usepackage{amsmath,amssymb}
\usepackage{bm,bbm}
\usepackage{graphicx, graphics, color}
\usepackage{slashed}
\usepackage{booktabs}
\usepackage{tabu}
\usepackage[dvipsnames]{xcolor}
\usepackage[normalem]{ulem}
\usepackage{tikz}
\usepackage{soul}
\usetikzlibrary{shapes,arrows}
\usetikzlibrary{positioning}
\usepackage[colorlinks=true,linkcolor=blue]{hyperref}

\newcommand{\tf}{{\rm T}}

\begin{document}

\title{First simultaneous extraction of spin-dependent parton distributions\\
	and fragmentation functions from a global QCD analysis}
\author{J. J. Ethier}
\affiliation{College of William and Mary, Williamsburg, Virginia 23187, USA}
\affiliation{Jefferson Lab, Newport News, Virginia 23606, USA}
\author{Nobuo Sato}
\affiliation{University of Connecticut, Storrs, Connecticut 06269, USA \\
\vspace*{0.2cm}
{\bf Jefferson Lab Angular Momentum (JAM) Collaboration
\vspace*{0.2cm} }}
\author{W. Melnitchouk}
\affiliation{Jefferson Lab, Newport News, Virginia 23606, USA}

\begin{abstract}
We perform the first global QCD analysis of polarized inclusive
and semi-inclusive deep-inelastic scattering and single-inclusive
$e^+e^-$ annihilation data, fitting simultaneously the parton
distribution and fragmentation functions using the iterative Monte
Carlo method. Without imposing SU(3) symmetry relations, we find
the strange polarization to be very small, consistent with zero
for both inclusive and semi-inclusive data, which provides a
resolution to the strange quark polarization puzzle.
The combined analysis also allows the direct extraction from data
of the isovector and octet axial charges, and is consistent with
a small SU(2) flavor asymmetry in the polarized sea.
\end{abstract}

\date{\today}
\maketitle


The decomposition of the proton's spin into its quark and gluon
helicity and orbital angular momentum contributions has been one of
the defining problems that has engaged the hadron physics community
for the better part of three decades \cite{Aidala12}.
Initial explanations of the small fraction of the proton spin found
to be carried by quarks focused on a large gluonic contribution
generated through the axial anomaly \cite{Altarelli88}, or a large
negative polarization of the strange quark sea.
Subsequent experiments failed to find compelling evidence to support
either of these scenarios, although recent results from RHIC have
provided the first clear indications for a nonzero gluon polarization,
$\Delta g$ \cite{DSSV14}.
Complementing this has been a growing effort to determine the quark
and gluon orbital angular momentum components of the proton spin,
through measurements of generalized parton distributions in exclusive
processes \cite{Diehl03}.
Critical to all these endeavors is the necessity to reliably extract
from the experimental data the fundamental parton distribution
functions (PDFs) that characterize the partons' spin and momentum
distributions through global QCD analysis.

Typically, global QCD analyses \cite{JAM15, LSS15, NNPDF14, BB10,
AAC09, DSSV09, KTA17} of inclusive deep-inelastic scattering (DIS)
and other polarized data extract spin-dependent PDFs using constraints
from weak baryon decays under the assumption of SU(3) flavor symmetry.
This puts significant restriction on the first moment of the
polarized strange PDF, with
	$\Delta s^+ \equiv \Delta s + \Delta \bar s
		    \approx -0.1$.
Further assumptions about the behavior of the PDFs at large parton
momentum fractions $x$ induces a shape for $\Delta s^+(x)$ with
magnitude peaking at $x \sim 0.1$.
With the inclusion of semi-inclusive DIS (SIDIS) data, a strikingly
different shape for the strange polarization emerges \cite{DSSV09,
LSS10}, changing sign to become positive at $x \sim 0.1$.  This was
found, however, to be strongly dependent on the assumed $s \to K$
fragmentation function (FF), which enters in the calculation of the
SIDIS cross section \cite{LSS10, LSS11}.  Ideally, an unambiguous
determination of the strange quark polarization requires a
{\it simultaneous} QCD analysis of both the PDFs and FFs.

In this paper we report on the first such analysis, using data from
inclusive and semi-inclusive DIS and single-inclusive $e^+ e^-$
annihilation (SIA) to simultaneously constrain the spin-dependent PDFs
and
$\pi^\pm$ and $K^\pm$ FFs.  To avoid biasing the extraction of
$\Delta s^+$ by assumptions about SU(3) symmetry, we allow for the
combined data sets to determine the octet axial charge directly.
This is not feasible in a DIS-only analysis, but becomes viable with
the flavor separation capability of SIDIS data.
We perform the analysis within the
iterative Monte Carlo (IMC) approach \cite{JAM15, JAM16},
which avoids potential bias in single-fit analyses introduced by
fixing parameters not well constrained by data, and allows a
statistically rigorous determination of PDF and FF uncertainties
by an efficient exploration of the parameter space.
%

In this first combined study of PDFs and FFs, which is performed
within collinear factorization at next-to-leading order (NLO) in the
$\overline{\rm MS}$ scheme, and referred to as ``JAM17'', we simplify
the analysis by placing cuts on the DIS and SIDIS kinematics to avoid
higher twist contributions, with the four-momentum transfer squared
	$Q^2 > 1$~GeV$^2$
and hadronic final state mass squared
	$W^2 > 10$~GeV$^2$.
The higher twists were extracted in a previous IMC
analysis~\cite{JAM15}, with a lower cut $W > 2$~GeV, but did not
significantly affect the determination of the leading twist PDFs.

The detailed expressions for DIS and SIA observables can be found
Refs.~\cite{JAM15} and \cite{JAM16}, respectively.
For the SIDIS data, the observables measured are the longitudinal
double spin asymmetries $A_1^h$ for the production of a hadron $h$,
\begin{equation}
A_1^h \left(x,z,Q^2\right) = \frac{g_1^h(x,z,Q^2)}{F_1(x,z,Q^2)},
\label{eq.A1h}
\end{equation}
where the semi-inclusive spin-dependent $g_1^h$ and spin-averaged
$F_1^h$ structure functions depend on both $x$ and the fraction
$z = p \cdot p_h/p \cdot q$ of the virtual photon's momentum ($q$)
carried by the 
hadron ($p_h$), with $p$ the target momentum.

The polarized $g_1^h$ function in Eq.~(\ref{eq.A1h}) is defined
in terms of the spin-dependent PDFs $\Delta q$ and FFs $D_q^h$,
\begin{eqnarray}
\label{eq.g1h}
\hspace*{-0.3cm}
g_1^h(x,z,Q^2)
&=& \frac{1}{2} \sum_q e_q^2
    \Delta q(x,Q^2) D_q^h(z,Q^2)
 +  {\cal O}(\alpha_s),
%
%
\end{eqnarray}
where the ${\cal O}(\alpha_s)$ corrections are given in
Ref.~\cite{Stratmann01}.
%
%
The unpolarized structure function $F_1^h$ is defined analogously,
with the spin-dependent PDFs 
replaced by their spin-averaged counterparts.

Following Refs.~\cite{JAM15, JAM16}, we parameterize both the
polarized PDFs and FFs at the input scale $Q_0^2=1$~GeV$^2$
using template functions of the form
\begin{align}
\tf(x;\bm{a})
= \frac{M\, x^a (1-x)^b (1 + c\sqrt{x})}
       {B(n+a,1+b) + c B(n+\frac12+a,1+b)},
\label{e.Txa}
\end{align}
where ${\bm a}=\{M,a,b,c\}$ are the fitting parameters,
and $B$ is the Euler beta function.
For the polarized PDFs we set $n=1$ so that $M$ corresponds to
the first moment.  This template is used for all the fitted
polarized PDFs, which we choose to be
	$\Delta q^+$,
	$\Delta \bar{q}$ and
	$\Delta g$,
for flavors $q = u, d$ and $s$.
The FFs are also given by Eq.~(\ref{e.Txa}) (with $x$ replaced by $z$),
setting $c=0$ and $n=2$, so that $M$ corresponds to the average
momentum fraction carried by the produced hadron.  For the FFs
	$D_{u^+}^{\pi^+} \equiv D_u^{\pi^+} + D_{\bar u}^{\pi^+}
			 = D_{d^+}^{\pi^+}$,
	$D_{u^+}^{K^+}$ and
	$D_{s^+}^{K^+}$,
which contain both favored and unfavored distributions,
we assign two template functions, while for the remaining
unfavored FFs,
	$D_{\bar u}^{\pi^+} = D_d^{\pi^+}$,
	$D_s^{\pi^+} = (1/2) D_{s^+}^{\pi^+}$,
	$D_{\bar u}^{K^+} = (1/2) D_{d^+}^{K^+}$ and
	$D_s^{K^+}$,
along with the heavy quarks and gluons, a single template function
is used.  Following Ref.~\cite{JAM16}, we use the zero mass variable
flavor scheme and parametrize the heavy quark FFs discontinuously
at their mass thresholds.

\begin{table}
\begin{tabular}{lcrr}
process                  & target       &$N_{\rm dat}$& $\chi^2$\ \ \\ \hline
DIS                      & $p,d,{}^3$He &  854     & 854.8	\\
SIA ($\pi^\pm$)       &	&  459    & 600.1 \\ 
SIA ($K^\pm$)         &	&  391     & 397.0     	\\
SIDIS ($\pi^\pm$)    	        &	&          & 	    	\\
 \ \ \ HERMES \cite{HERMES05}   & $d$   &   18     &  28.1   	\\
 \ \ \ HERMES \cite{HERMES05}   & $p$   &   18     &  14.2   	\\
 \ \ \ COMPASS \cite{COMPASS09} & $d$   &   20     &   8.0    	\\ 
 \ \ \ COMPASS \cite{COMPASS10} & $p$   &   24     &  18.2	\\
SIDIS ($K^\pm$)                 &       &          & 	        \\
 \ \ \ HERMES \cite{HERMES05}   & $d$   &   27     &  18.3	\\
 \ \ \ COMPASS \cite{COMPASS09} & $d$   &   20     &  18.7	\\  
 \ \ \ COMPASS \cite{COMPASS10} & $p$   &   24     &  12.3  	\\ \hline
{\bf Total:}          		&       &\ \ {\bf 1855}&\ \ {\bf 1969.7}\\
\end{tabular}
\caption{Summary of $\chi^2$ values and number of data points
	$N_{\rm dat}$ for the various processes used in this
	analysis.}
\label{t.chi2}
\end{table}

The resulting $\chi^2$ values for each process fitted in our analysis
are presented in Table~\ref{t.chi2}.
For inclusive DIS we use the data sets from
  Refs.~\cite{EMC89, SMC98, SMC99, COMPASSdis10, COMPASSdis07,
  SLAC-E130, SLAC-E142, SLAC-E143, SLAC-E154, SLAC-E155p,
  SLAC-E155d, SLAC-E155_A2pd, SLAC-E155x,
  HERMES97, HERMES07, HERMES12},
and for SIA from
  Refs.~\cite{TASSO80, TASSO83, TASSO89, ARGUS89, TPC84, TPC86, TPC88,
  HRS87, SLD04, OPAL94, OPAL00, ALEPH95, DELPHI95, DELPHI98, TOPAZ95,
  Belle13, Leitgab13, BaBar13}.
%
%
The SIDIS data sets are from HERMES~\cite{HERMES05}
for $\pi^\pm$ and $K^\pm$ production from the deuteron, and $\pi^\pm$
production from the proton, and from COMPASS with
$\pi^\pm$ and $K^\pm$ production from deuterium~\cite{COMPASS09}
and hydrogen~\cite{COMPASS10} targets.
Overall, the $\chi^2$ per datum for all the SIDIS $\pi^\pm$ data is
68.5/80, and 49.3/71 for the $K^\pm$ data, while the $\chi^2$ per datum
for the combined inclusive DIS, SIDIS and SIA data is
	$1969.7/1855 \approx 1.06$.

\begin{figure}[t]
\includegraphics[width=0.48\textwidth]{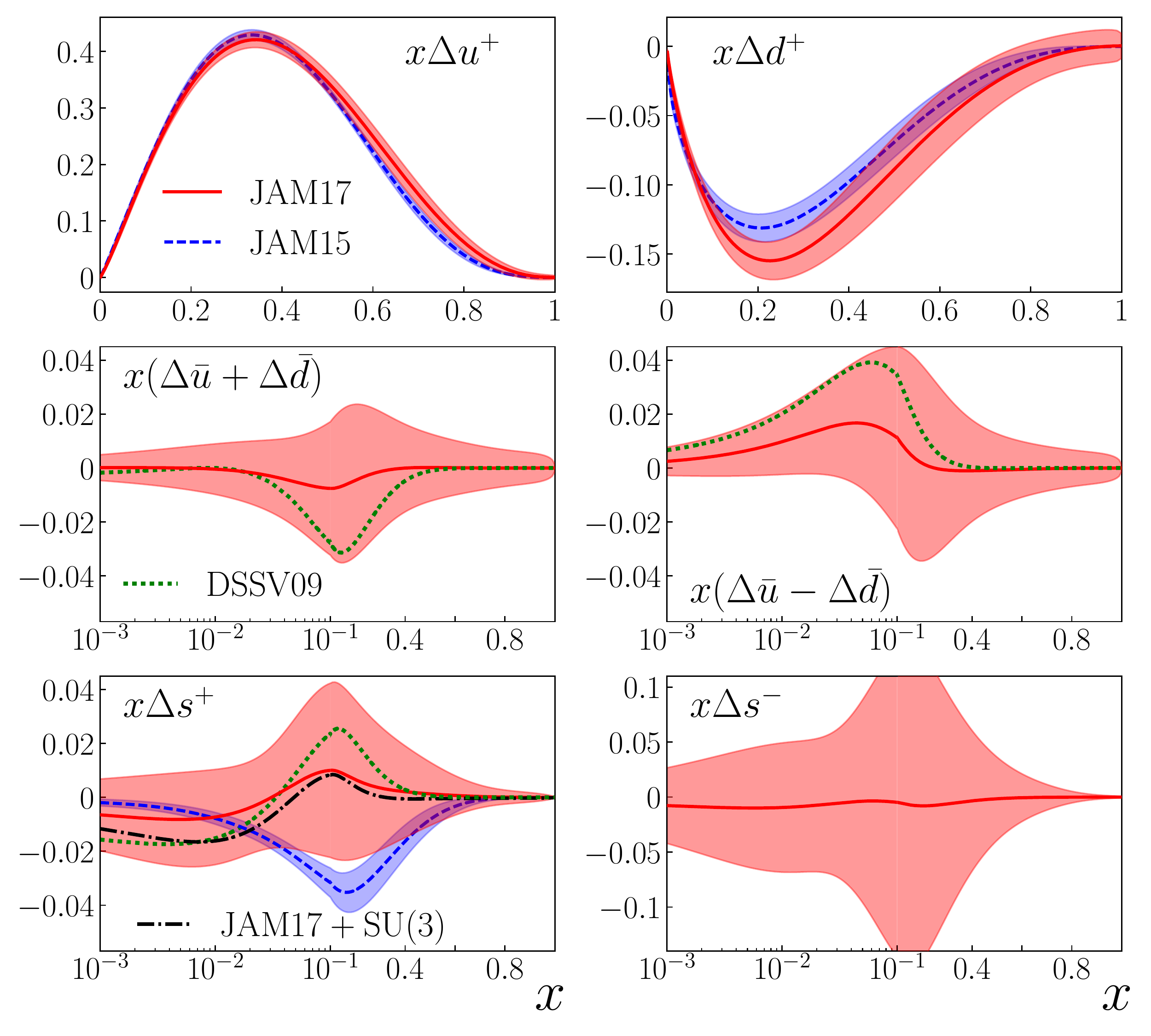}
\caption{Spin-dependent PDFs with 1$\sigma$ uncertainty bands
	from the JAM17 fit at the input scale $Q_0^2 = 1$~GeV$^2$.
	The full results (red solid curves) are compared with the
	JAM15 $\Delta q^+$ PDFs \cite{JAM15} (blue dashed curves)
	and with the DSSV09 fit \cite{DSSV09} for sea quark PDFs
	(green dotted curves).
	The $\Delta s^+$ PDF is also compared with the JAM17 fit
	including the SU(3) constraint on the octet axial charge
	(black dot-dashed curve).}
\label{f.ppdfQ20} 
\end{figure}

The polarized quark and antiquark PDFs from the combined fit are
illustrated in Fig.~\ref{f.ppdfQ20}, together with their 1$\sigma$
uncertainties.
(The polarized gluon PDF is essentially unchanged from the
earlier JAM15 analysis \cite{JAM15}.)
For the denominator of the asymmetries $A_1^h$, we use spin-averaged
PDFs from the CJ12 NLO global fit \cite{CJ12}.
Using the MMHT14 \cite{MMHT14} PDFs instead gives a difference of
$\approx 2\%-5\%$, which is insignificant on the scale of the
experimental uncertainties of the asymmetries.
The $\Delta u^+$ and $\Delta d^+$ PDFs, which are determined largely
by the inclusive DIS data, are similar to those in the JAM15
analysis~\cite{JAM15}, giving only marginally harder distributions at
large $x$ values.  The difference in the magnitudes of $\Delta d^+$
at $x \sim 0.2$ arises from anticorrelation with $\Delta s^+$;
since the latter is less negative, it requires some compensation
to describe the DIS observables.

Unlike inclusive DIS, the SIDIS observables can in principle
discriminate between different quark and antiquark flavors,
and in Fig.~\ref{f.ppdfQ20} we also show the light sea quark
polarizations for the isoscalar and isovector combinations,
$\Delta \bar u \pm \Delta \bar d$.
Our results suggest a slightly positive isovector sea polarization
in the range $x \approx 0.01 - 0.1$, with the isoscalar
combination more consistent with zero.
This is similar to the expectations in some nonperturbative models
\cite{Schreiber91, Diakonov96} that predict larger isovector than
isoscalar sea polarization,
as well as in recent lattice simulations~\cite{HueyWen, Fernanda}.
The signal is relatively weak, however, and can be attributed to
several $\pi^\pm$ and $K^\pm$ SIDIS data sets that marginally favor
a nonzero sea polarization.

\begin{figure}[t]
\includegraphics[width=0.49\textwidth]{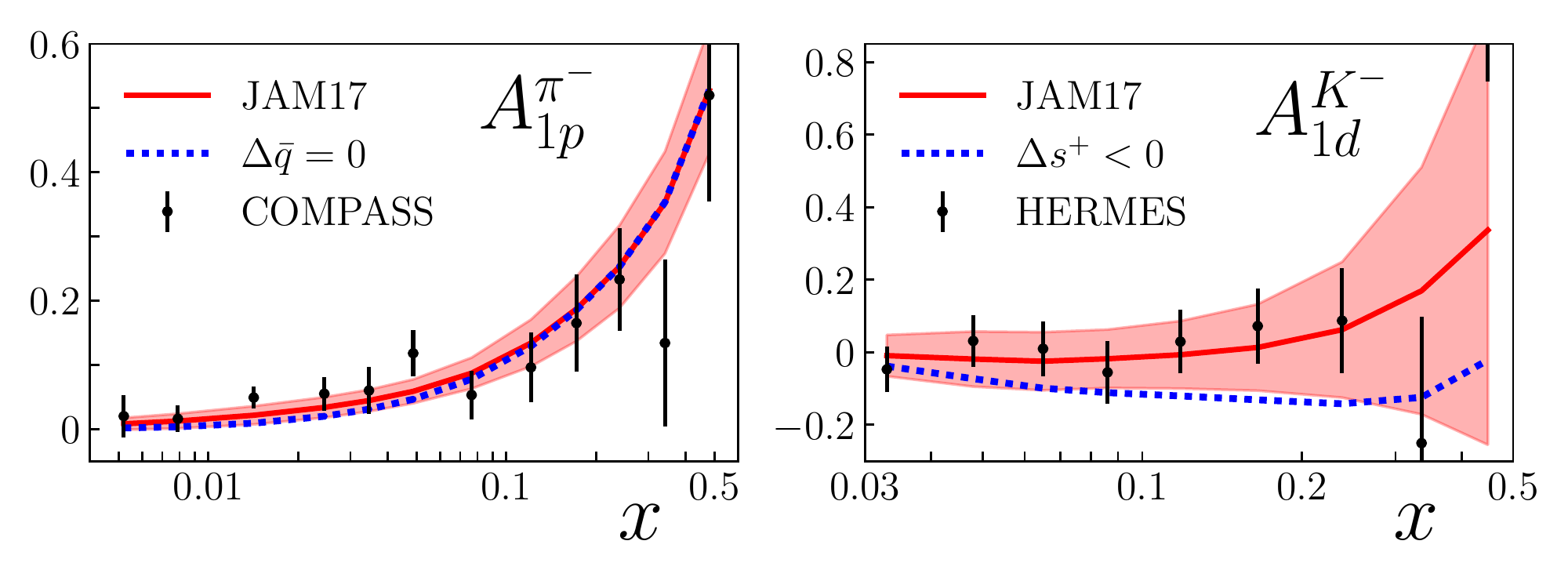}
\caption{Semi-inclusive polarization asymmetries
	$A_{1p}^{\pi^-}$ from COMPASS \cite{COMPASS10} (left) and
	$A_{1d}^{K^-}$ from HERMES \cite{HERMES05} (right)
	compared with the full JAM17 fit (red curves and band)
	and with the result assuming
	$\Delta \bar q \equiv \Delta \bar u = \Delta \bar d = 0$
		(for $A_{1p}^{\pi^-}$)
	and the (negative) $\Delta s^+$ from JAM15 \cite{JAM15}
		(for $A_{1d}^{K^-}$).}
\label{f.DvT} 
\end{figure}

An example of this is illustrated in Fig.~\ref{f.DvT} for the
\mbox{COMPASS} $\pi^-$ asymmetry \cite{COMPASS10}, which,
because of the valence ($\bar{u} d$) structure of the $\pi^-$,
is the most sensitive observable to $\bar u$ polarization.
Comparing the fitted proton $A_{1p}^{\pi^-}$ asymmetry with
that obtained by setting $\Delta \bar{u} = \Delta \bar{d} = 0$,
the difference is rather small, but noticeable for $x \lesssim 0.1$,
where the asymmetry with the unpolarized sea lies at the edge of
the 1$\sigma$ envelope of the full result.
Similar effects are found for other SIDIS asymmetries that
depend explicitly on $\Delta \bar u$ or $\Delta \bar d$.
The results are also qualitatively similar to those found in the
DSSV09 global analysis \cite{DSSV09}, although the magnitude of the
sea quark asymmetries here is somewhat smaller.
%

For the strange quark polarization, the results in Fig.~\ref{f.ppdfQ20}
suggest that $\Delta s^+$ is small at all $x$, albeit within
relatively large uncertainties.
While consistent with zero within 1$\sigma$, there does
appear some indication of a positive $\Delta s^+$ at $x \approx 0.1$.
This can be attributed directly to the HERMES deuteron $K^-$
production data~\cite{HERMES05}, illustrated in Fig.~\ref{f.DvT}.
Since $\Delta s$ is weighted by the (large) favored $D_s^{K^-}$ FF,
the $A_{1d}^{K^-}$ asymmetry is most sensitive to strange quark
polarization.
In contrast, $K^+$ production, which is sensitive to $\Delta \bar s$,
is dominated by the much larger $\Delta u$ PDF weighted by the favored
$D_u^{K^+}$.

Compared with the full result, the asymmetry computed with a negative
$\Delta s^+$, as in the JAM15 analysis of inclusive DIS \cite{JAM15},
gives a significantly worse fit to the HERMES $A_{1d}^{K^-}$ data,
with $\chi^2$ increasing
	from 5.7 to 18.5 for 9 data points.
A similar effect is seen for the COMPASS $K^-$ data on protons
(deuterons), which prefer a non-negative strangeness, with $\chi^2$
increasing
	from 4.8 to 9.0 (12.0 to 18.5) for 12 (10) data points.

In addition to $\Delta s^+$, we also explored the sensitivity
to a nonzero strange--antistrange asymmetry, $\Delta s^-$.
While most global PDF analyses assume $\Delta s = \Delta \bar s$,
a nonzero asymmetry is expected from chiral symmetry breaking
in QCD~\cite{Signal87, Malheiro97, Thomas00}.
In principle, the availability of precise $K^\pm$ SIDIS data
could discriminate between $s$ and $\bar s$ polarization; however,
as Fig.~\ref{f.ppdfQ20} illustrates, the current experimental errors
render extraction of a nonzero $\Delta s^-$ signal impractical.

\begin{figure}[t]
\includegraphics[width=0.48\textwidth]{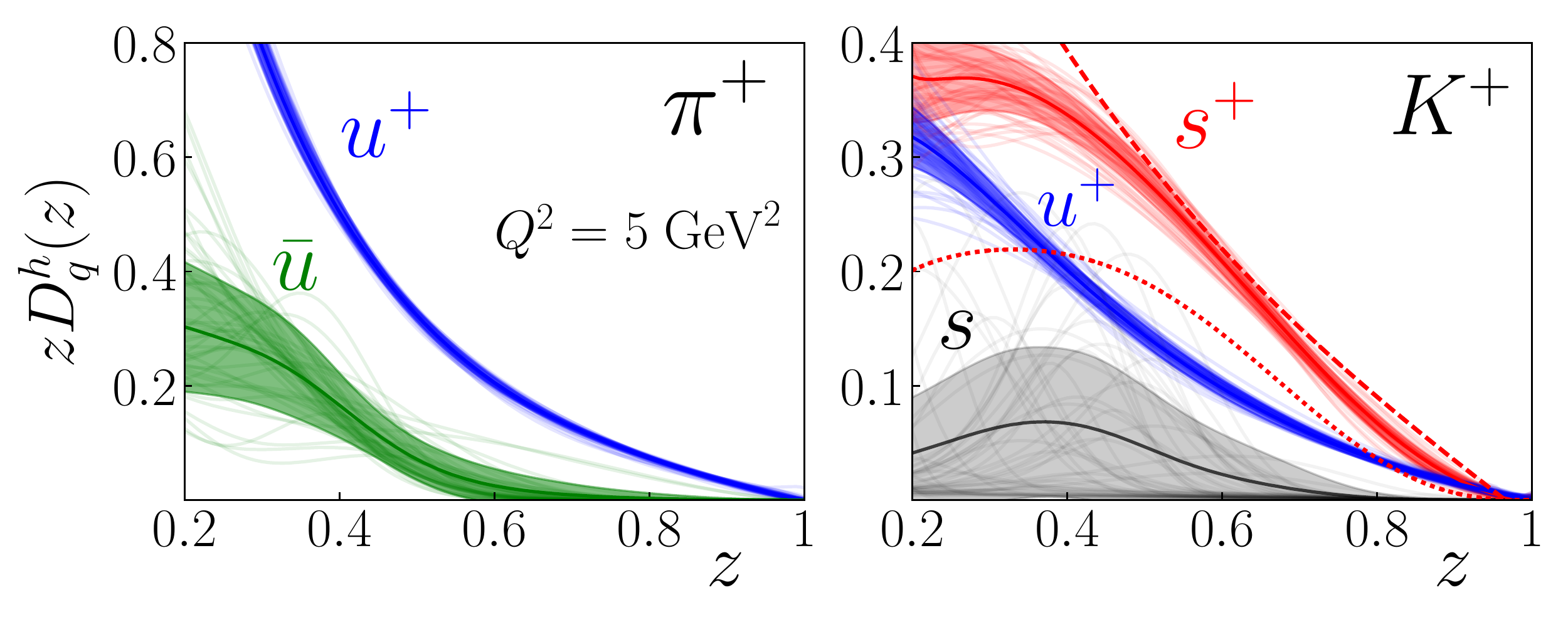}
\caption{Fragmentation functions $z D_q^h$ to
	$\pi^+$ (left panel) and $K^+$ (right panel) for
	$u^+$ (blue), $\bar u$ (green),	$s^+$ (red) and $s$ (grey)
	at $Q^2 = 5$~GeV$^2$ for the JAM17 analysis.
	Random samples of 50 posteriors are shown with the
	mean and variance, and compared with the $s^+ \to K^+$ FFs
	from DSS~\cite{DSS} (dashed) and HKNS \cite{HKNS} (dotted).}
\label{f.FFs} 
\end{figure}

Of course, preference for a positive or negative strangeness depends
rather strongly on the FFs used in the evaluation of the asymmetry
\cite{LSS10, LSS11, LSS15}.
The solution to this problem is to simultaneously determine both PDFs
and FFs, as we seek to do here.
The results for the FFs extracted from the combined fit are displayed
in Fig.~\ref{f.FFs} for the most relevant quark flavors fragmenting
to $\pi^+$ and $K^+$, at a scale $Q^2 = 5$~GeV$^2$ appropriate for
the SIDIS data.

For the pion, the $D_{u^+}^{\pi^+}$ FF is relatively well constrained
by the SIA data, compared with the unfavored $D_{\bar u}^{\pi^+}$.
The $\pi^+$ FFs are also similar to those from the previous JAM
analysis of SIA data \cite{JAM16}, as well as from other
parametrizations \cite{DSS, HKNS}. 
For kaons, the uncertainties for the
favored $D_{u^+}^{K^+}$ and unfavored $D_s^{K^+}$ are generally larger
because of the lower precision of the $K$
data.

One of the most important features in Fig.~\ref{f.FFs} is the
difference between the $D_{s^+}^{K^+}$ FF for the various
parametrizations, which has profound impact on $\Delta s$
extraction.  Here the JAM17 result is more comparable with the
DSS fit~\cite{DSS}, while the magnitude of the HKNS~\cite{HKNS}
result is somewhat smaller.
The $D_{s^+}^{K^+}$ FF is also qualitatively similar to the
recent NJL--Jet model calculation \cite{Hrayr11}, specifically
in the large-$z$ region where $D_{s^+}^{K^+} > D_{u^+}^{K^+}$.
The $D_{s^+}^{K^+}$ obtained from the SIA-only analysis \cite{JAM16}
is very similar to that in Fig.~\ref{f.FFs}, with the SIDIS data
pulling the JAM17 result slightly larger at low $z$.
Recall that the smaller $s^+ \to K^+$ fragmentation in the HKNS fit
is what allowed a more negative $\Delta s^+$ at $x \sim 0.1$ in the
combined DIS and SIDIS analysis of Ref.~\cite{LSS11}, similar to
the shape of $\Delta s^+$ in DIS-only analyses such as JAM15 in
Fig.~\ref{f.ppdfQ20}.  The shapes of the heavy quark and gluon FFs
are essentially unchanged from Ref.~\cite{JAM16}.

If it is SIDIS data that restrict the $\Delta s^+(x)$ PDF to be small
and positive at intermediate $x$ values, a natural question to ask is
what drives $\Delta s^+$ to be large and negative in DIS-only analyses?
The answer appears to be the imposition of the SU(3) constraint on the
octet axial charge, which is related to the lowest moment of the SU(3)
nonsinglet combination,
	$a_8 \equiv \Delta u^+ + \Delta d^+ - 2 \Delta s^+$.
We have verified that in the absence of this constraint, it is indeed
possible to fit the DIS data sets with zero $\Delta s^+$ at the input
scale, with identical $\chi^2$ values as in a fit with the strange
distribution free to vary.  This confirms that the sensitivity to
$\Delta s^+$ from DIS data alone, in combination with $Q^2$ evolution,
is negligible.  In contrast, the SU(3) assumption tends to pull the
strange PDF to be negative across all $x$ in order to generate a
negative moment, $\Delta s^+(Q_0^2) \approx -0.1$.

To understand the origin of the large negative peak in $\Delta s^+$
at $x \approx 0.1$ in DIS-only analyses, we examine the behavior of
$\Delta s^+$ in the absence of low-$x$ DIS data ($x \lesssim 0.02$),
where sea quarks are expected to play a greater role.
Starting from a shape of $\Delta s^+(x)$ at the input scale that is
negative at small $x$ and positive at large $x$, such as in the DSSV09
fit~\cite{DSSV09}, we find that the strange PDF remains qualitatively
unchanged when fitting to the reduced DIS data set.
Upon closer examination, around 5 data points at the lowest $x$ bins
from the COMPASS deuterium data are found to favor small negative
values for $\Delta s^+$, which then drives the strange PDF in the
intermediate-$x$ region to be more negative in order to satisfy the
SU(3) constraint.
Finally, the characteristic negative peak at $x \sim 0.1$ observed
in most global analyses is generated by fixing the parameter
$b \approx 6-10$, as for typical sea quark PDFs.
Such a peak is artificial since there is no direct sensitivity to
$\Delta s^+(x)$ in current inclusive DIS data.

The ``strange quark polarization puzzle'' \cite{LSS11, LSS15} can
therefore be understood by simply relaxing the SU(3) constraint,
which then produces a strange distribution with shape and magnitude
that agree well with DIS and SIDIS asymmetries.  While both the
(mostly positive) JAM17 and (mostly negative) JAM15 strange PDFs
give nearly identical $\chi^2$ for DIS data, the latter will be
strongly disfavored by the SIDIS asymmetries.
In fact, the positive shape of $\Delta s^+$ at $x \sim 0.1$
is obtained even when samples consistent with SU(3) symmetry
are selected, as Fig.~\ref{f.ppdfQ20} illustrates. 
Such samples prefer more negative PDFs at lower $x$ values,
$x \lesssim 10^{-2}$, where the shape is not well constrained,
and restrict the first moment to $\Delta s^+ \sim -0.1$.
In contrast, our new results gives a smaller averaged value,
	$\Delta s^+(Q_0^2) = -0.03(10)$,
but now of course with larger uncertainty.
Interestingly, the central value agrees with the recent lattice QCD
determination of strangeness polarization,
	$\Delta s^+_{\rm latt} = -0.02(1)$
at $Q^2 \approx 7$~GeV$^2$~\cite{Bali12}.

\begin{figure}[t]
\includegraphics[width=0.48\textwidth]{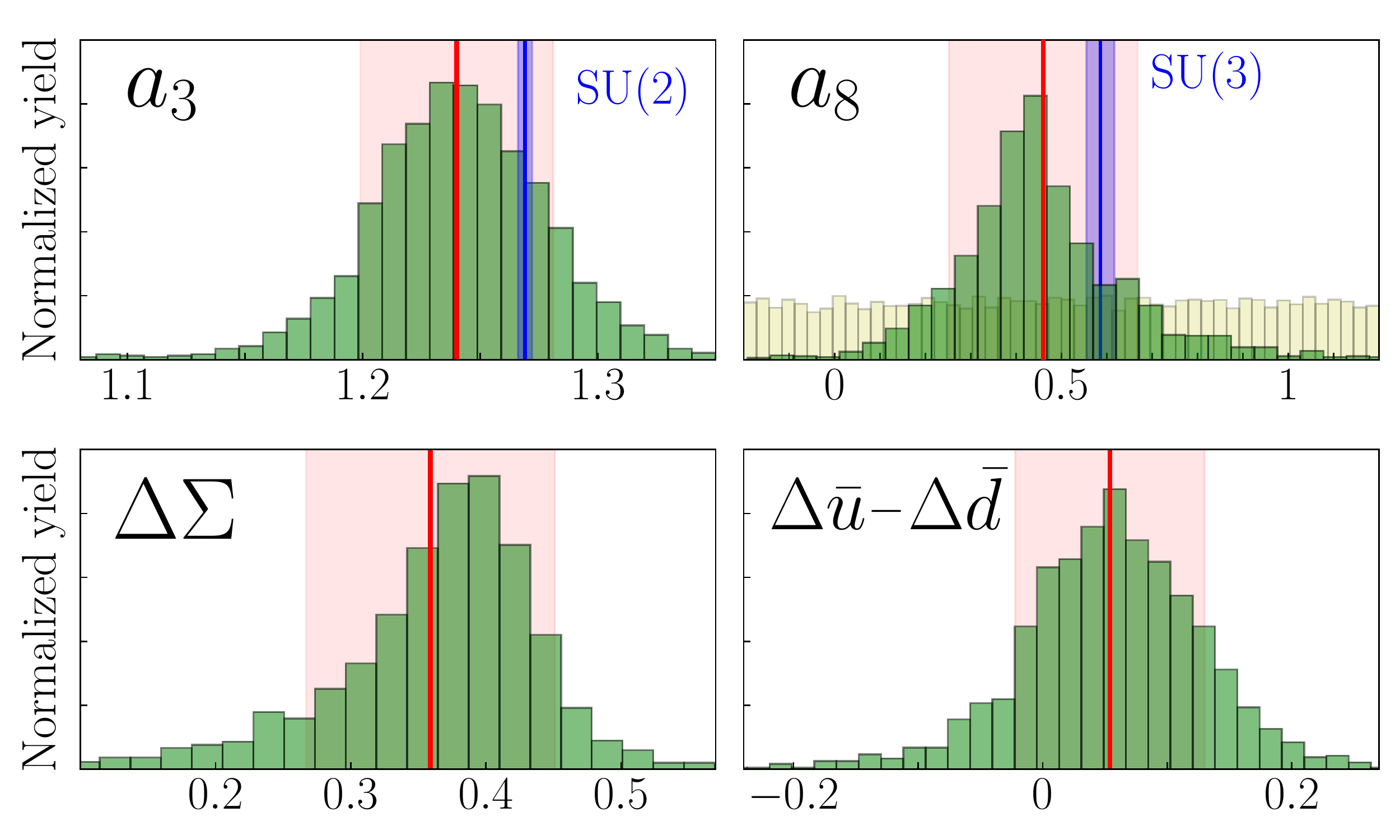}
\caption{Normalized yield of the lowest moments of the
	spin-dependent PDFs for the triplet ($a_3$),
	octet ($a_8$) and singlet ($\Delta \Sigma$)
	axial charges, and the flavor asymmetry
	$\Delta \bar u - \Delta \bar d$,
	with average values (red vertical lines) and 1$\sigma$
	deviations (pink bands) indicated at the input scale.
	For the scale invariant $a_3$ [$a_8$], the SU(2) [SU(3)]
	symmetric values are indicated (blue vertical bands),
	together with the flat prior distributions for $a_8$
	without SU(3) (yellow histograms).}
\label{f.moments} 
\end{figure}

Our result for the strange moment translates to a central value of the
octet axial charge $a_8 = 0.46(21)$ that is $\approx 20\%$ smaller
than the traditional SU(3) value $0.586(31)$,
as suggested in earlier theoretical studies \cite{Bass10}.
Even though the uncertainty is somewhat large, the peaking of the
$a_8$ distribution around $\sim 0.5$ is entirely data driven,
as Fig.~\ref{f.moments} illustrates with the comparison of the
flat prior distributions sampled in the range $[-0.2,1.2]$.
Future higher precision SIDIS kaon data would be needed to reduce
the uncertainty on both the polarized strangeness and test the degree
of SU(3) breaking in the octet axial charge.

Another consequence of the more positive value of $\Delta s^+$
(smaller $a_8$) is an $\approx 25\%$ larger total spin carried by
quarks and antiquarks in the nucleon~\cite{Bass10},
	$\Delta \Sigma(Q_0^2) = 0.36(9)$.
Within the larger uncertainties resulting from the relaxing of the
SU(3) constraint in our simultaneous analysis, this is compatible with
the singlet charge of 0.28(4) obtained in the JAM15 fit \cite{JAM15}.
In fact, the simultaneous fit can also be used to determine the
triplet axial charge $a_3 \equiv \Delta u^+ - \Delta d^+$
preferred by the data, without assuming SU(2) symmetry.
We find that
	$a_3 = 1.24(4)$,
in good agreement with the standard value $g_A = 1.269(3)$
from neutron weak decay.
This is a remarkable empirical confirmation of the equality
between $a_3$ and $g_A$, and of QCD itself, to almost 2\%!

Finally, as suggested in Fig.~\ref{f.ppdfQ20}, the antiquark component
of the isovector axial charge prefers slightly positive values,
	$\Delta \bar u - \Delta \bar d = 0.05(8)$,
but is consistent with zero within the uncertainty.
The recent polarized $pp$ scattering data from PHENIX \cite{PHENIX-W}
on asymmetries from $W^\pm + Z$ decays and from STAR \cite{STAR-W} on
$W^\pm$ asymmetries also indicate a slightly larger $\Delta \bar u$
in the $x \sim 0.16$ range.
%

In the future, the IMC analysis will be extended to include NNLO
\cite{Anderle:2015lqa,Bertone:2017tyb} and small-$x$
\cite{Kovchegov:2017jxc, Kovchegov:2017lsr} corrections,
as well as unpolarized SIDIS data for better determination
of the unfavored FFs \cite{DSSpi15, DSSk17}.
Since the unpolarized strange quark PDF is currently not well
determined, useful constraints on $\Delta s$ from these data
will necessitate a simultaneous analysis of spin-averaged and
spin-dependent PDFs, in addition to FFs.
This remains an important challenge for future global QCD analyses.

We are grateful to A.~Accardi, S.~D.~Bass and A.~W.~Thomas for
helpful discussions.
This work was supported by the US Department of Energy (DOE) contract
No.~DE-AC05-06OR23177, under which Jefferson Science Associates, LLC
operates Jefferson Lab. N.S. was supported by the DOE contract
DE-FG-04ER41309.


\end{document}